\documentclass[a4paper, 11pt]{article}
\usepackage[dvips]{color}

\newcommand{\blue}{\color{blue}}

\renewcommand{\v}[1]{\vspace{#1 cm}}

 \newcommand{\Ohm}{$\Omega\;$}

\newcommand{\beq}{\begin{equation}} \newcommand{\eeq}{\end{equation}}
\newcommand{\beqa}{\begin{eqnarray}} \newcommand{\eeqa}{\end{eqnarray}}

\newcommand{\newslide}[1]{\begin{slide}\begin{center}{\blue
          #1} \end{center} \small}
\newcommand{\myendofslide}[1]{\vspace{\fill} \tiny \begin{flushright} 
          #1 \end{flushright} \end{slide} }

\usepackage{epsfig} 
\usepackage{makeidx} \makeindex

\usepackage{graphicx}
\begin{document}

%\hspace{\fill} {\large LHCb internal note -- inner tracking -- LHCb 2000-056}

%\hspace{\fill} {\large \today}

\begin{center}

%{\Huge PRELIMINARY}

\v{1}

{\huge \bf A triple GEM detector\\ with two dimensional readout}

\vspace{1ex}

%{\Large PSI test beam results obtained in December 1999}

\vspace{2ex}

M. Ziegler, P. Sievers, U. Straumann\\[2ex]
  {\it Physik Institut Universit\"at Z\"{u}rich}\\[2ex]
\today

\vspace{3ex}
\end{center}

{\tt This is a reduced version for hep-ex, from which the large photographs
have been omitted to keep it small in size. A full version is available from:\\
http://www.physik.unizh.ch/groups/groupstraumann/papers/main.ps.gz}
\vspace{3ex}

{\setlength{\parskip}{0cm}\tableofcontents}

\vspace{3ex}

The triple GEM detector is a micropattern gas detector which consists of a
primary ionisation gap and three consecutive gas electron multiplier (GEM)
foils \cite{GEM}. A printed circuit board with readout strips detects the
current induced by the drifting electron cloud originating from the last GEM
stage. Thus the gas amplification and the signal readout are completely
separated.  Triple GEM detectors are being developed as a possible technology
for the inner tracking in the LHCb experiment.

In an earlier note we have reported first experience with such a detector in a
test beam at PSI\footnote{Paul Scherrer Institut, Switzerland.
  http://www.psi.ch} \cite{triplegem1}. Here we describe the construction of
an improved version (thinner transfer gaps, segmented GEM foils, two
dimensional readout). Results from performance measurements are presented
using intense hadronic beams as well as cosmic ray data.

\section{Requirements on an inner tracking detector for LHCb}

LHCb \cite{proposal} is a second generation experiment on $b$ quark physics
which will be operated at the LHC (Large Hadron Collider) at CERN. The goal of
the experiment is to measure systematically all observable CP violation
effects and the rare decays in the $B$-meson system with unprecedented
precision. It will allow to improve the knowledge on standard model physics
and look for new CP violation effects possibly induced from physics beyond the
standard model.

The analysis of rare $B$ decay channels requires an excellent resolution of
the kinematic quantities of the decay products, their momentum, invariant
masses and vertex position. A beam axial magnetic spectrometer with a large,
high resolution tracking system complemented with particle identification
devices (RICH, electromagnetic and hadronic calorimeters and a muon system)
provides optimal performance for $B$ physics studies at LHC.

The tracking system consists of three parts: A vertex detector (silicon
microstrip technology within the vacuum chamber of LHC), an outer tracking
system to be installed in areas with small particle occupancies and an
inner tracking system for areas with large particle occupancies. For the
latter the collaboration recently decided among the various micropattern gas
detector technologies to choose the triple GEM detector as described
in this note. However this gas detector is still in competition with a
microstrip silicon detector, the decision for one or the other of the two is
presently foreseen to be made by end of the year 2001.

There are 11 inner tracking stations mounted vertically around the beam axis,
each with a size of 60 cm (horizontal) $\times$ 40 cm (vertical). Momentum and
invariant mass resolution ask for a position resolution of $\sigma < 200
\mu$m, while the radiation length should be kept as small as possible (in the
simulations so far a value of about 2\% per station is assumed). A stereo
angle of 100 mrad is foreseen and each station needs to provide 4
measurements: $x$, $u$, $v$, $x$, where $u$ and $v$ denote the stereo layers.
For an efficient pattern recognition the channel occupancy has to be kept below
a few \%, which also requires the time occupancy of the signals not to exceed
about two or three LHC bunch crossings of 25 ns.

Simulation of the LHCb interaction region, including the beam vacuum chamber,
were performed to estimate the particle fluxes \cite{talanov}, using an
average nominal luminosity of 2$\times$10$^{32}$~cm$^{-2}$s$^{-1}$. The peak
luminosity is assumed to be higher by a factor 2.5. The particle fluxes and
its compositions depend strongly on the position parallel to the beam axis $z$
and on the distance from the beam: the maximum charged hadron rate is expected
to be 8$\times$10$^3$ mm$^{-2}$s$^{-1}$ and the maximum electron and positron
rate (originating mostly from photon interaction with the beam pipe) might be
as high as 5$\times$10$^4$~mm$^{-2}$s$^{-1}$. A large fraction of the latter
enters the detector under a large inclination angle. The total counting rate
per station reaches between 200 and 1000 MHz, again depending on $z$. These
rates correspond to a maximum total radiation dose of about 1.6 Mrad/year.

A major problem of gas micropattern detectors comprises the fact, that in
hadronic particle beams from time to time very high primary ionisations occur,
which may lead to high voltage breakdowns producing dead time, huge
electronic noise to other detectors and even may damage the detector and the
readout electronics. For GEM foils it has been shown, that they usually
withstand several hundred sparks per cm$^{2}$ without permanent damage
\cite{perroud}. We require both, that the spark rate is limited
to one per hour per tracking plane and the detector should stay
alive for several years. Taking into account the expected hadron 
particle flux it follows, that the discharge probability 
per incident particle must not exceed 10$^{-12}$.

\section{Detector construction}

\begin{figure}
\begin{center}
\epsfig{file=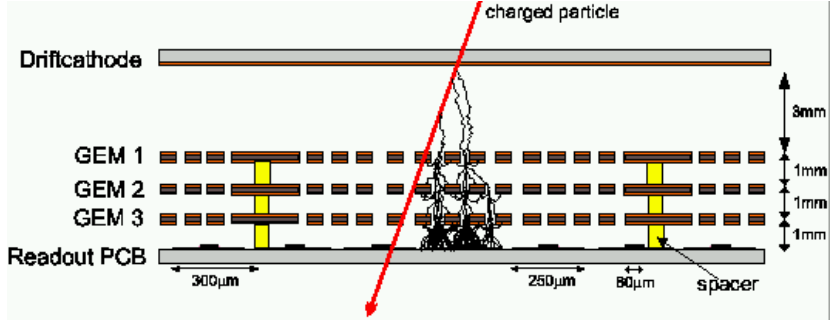,width=0.7\linewidth}
\end{center}

\caption{Schematic view of the detector with three GEM foils, 
  which are seperated by 1 mm gaps using spacers. Pitch and width
of the strips on the two coordinate readout board are also shown.}
\label{TripleGEM}
\end{figure}

The triple GEM detector is depicted schematically in figure \ref{TripleGEM}.
The size of the active surface is 30 $\times$ 23 cm$^2$.

\subsection{Drift gap and amplification stages}

\begin{figure}
\begin{minipage}[t]{.48\textwidth}
\begin{center}
\caption{GEM foil with a
    200 $\mu$m wide gap between segments, where the copper is removed.}
\label{segment}
\end{center}
\end{minipage}\hfill
\begin{minipage}[t]{.48\textwidth}
\begin{center}
\caption{A single spacer
  glued onto the GEM foil onto a hole free area.} \label{spacer}
\end{center}
\end{minipage}
\end{figure}

The drift cathode consists of a 50 $\mu$m thick Kapton foil covered
with a 15 $\mu$m thick copper layer. The drift gap is 3 mm wide.

The gas amplification takes place in three stages in the holes of the GEM
foils\footnote{The GEM foils were manufactured by Rui de Oliveira et al. in the CERN workshop},
which are mounted with a gap of 1 mm between them.  The basis material
for the foils is 50~$\mu$m thick Kapton with a 15~$\mu$m thick copper cladding
on both sides.  The hole parameters are chosen in accordance with results of
studies made by the CERN GDD group \cite{bachmann}. The holes are arranged
with a pitch of 140~$\mu$m, the diameter of the copper hole is 80~$\mu$m and
the diameter of the hole in the Kapton is 50~$\mu$m (see also figure \ref{segment}).

The capacity between the two electrodes of a GEM foil is as large as 30 nF. To
reduce the total charge released in a high voltage breakdown the GEM electrode
is divided on one side into 10 rectangular segments, which are separated by
copper free gaps of 200~$\mu$m width. Within these gaps there are no GEM holes
(figure \ref{segment}). Each segment is powered through an individual HV
resistor of 10 M\Ohm and 1 M\Ohm resistor for the first two and the
last GEM foil respectively outside the detector. This segmentation has the
additional advantage, that only a small fraction of the detector would fail in
the unlikely case of a high voltage breakdown causing a permanent GEM short.

\begin{figure}
\begin{minipage}[t]{.48\textwidth}
\begin{center}
\caption{A GEM foil
  stretched with tape to a large auxiliary aluminum frame during glueing.}
\label{gem-ganz}

\end{center}
\end{minipage}\hfill
\begin{minipage}[t]{.48\textwidth}
\begin{center}
\caption{HV
    resistor connected to GEM segment, 1mm G10 frame and spacer.}
  \label{rahmen-segment}
\end{center}
\end{minipage}
\end{figure}

To avoid contact between two GEM foils due to electrostatic forces or other
mechanical instabilities spacers are glued onto them about every 4.5 cm.
Since HV breakdown problems may occur due to unintentionally deposited glue in
the GEM holes, the GEM contains 2 mm diameter regions without holes, where
spacers can be placed (see figure \ref{spacer}). The spacers consist of small
cylinders with diameter 1.1 mm and height 1 mm of pure epoxy resin
H72\footnote{EPO--TEK H72, Polytek GmbH, D-76337 Waldbronn, Germany}. They
are glued with the same epoxy resin onto the GEM foil. Each GEM is equipped
with 35 spacers, which were positioned by hand under a microscope. This
procedure took about 45 minutes per foil. The total fraction of inactive surface
of the detector due to spacers is $1.6 \times 10^{-3}$, however the
inefficency introduced is expected to be even smaller, since typical charge
cluster diameters resulting from particle tracks are of order 0.5 mm.

The detector was mounted together starting with the drift cathode. An aluminum
frame was glued mit the same H72 glue on the Kapton side and a 3 mm G10 frame
onto the copper side simultanously. The aluminum frame is required to assure
mechanical stability of the detector. In future this frame should be replaced
by a full size plate of thin material (e.g. Rohacell) to reduce the total
radiation length.

Then the GEM foil was stretched by putting it onto a large aluminum frame and
pulling it with tape to the sides (figure \ref{gem-ganz}) with the spacers
already in place. Now a high voltage test in a nitrogen atmosphere was
performed.  The voltage applied to the foils was increased continously during
about two hours until sparking started somewhere between 580V and 650V.  The
GEMs were kept sparking about twenty times before decreasing the voltage
again to make sure that dust possibly trapped in the GEM holes would burn out
without generating a short circuit between the GEM surfaces.

After this test was successfully completed the first GEM foil was
simultanously glued to the 3 mm G10 frame and the first 1 mm frame onto the
other side of this GEM.  Afterwards the next two GEMs were stretched, tested
and glued in the same way. An overview of a part of the detector in this state
is shown in figure \ref{rahmen-segment}.  Finally the hole sandwich
structure was glued to the readout board.

\subsection{High voltage supply}

\begin{figure}
\centering
\includegraphics[width=0.7\textwidth]{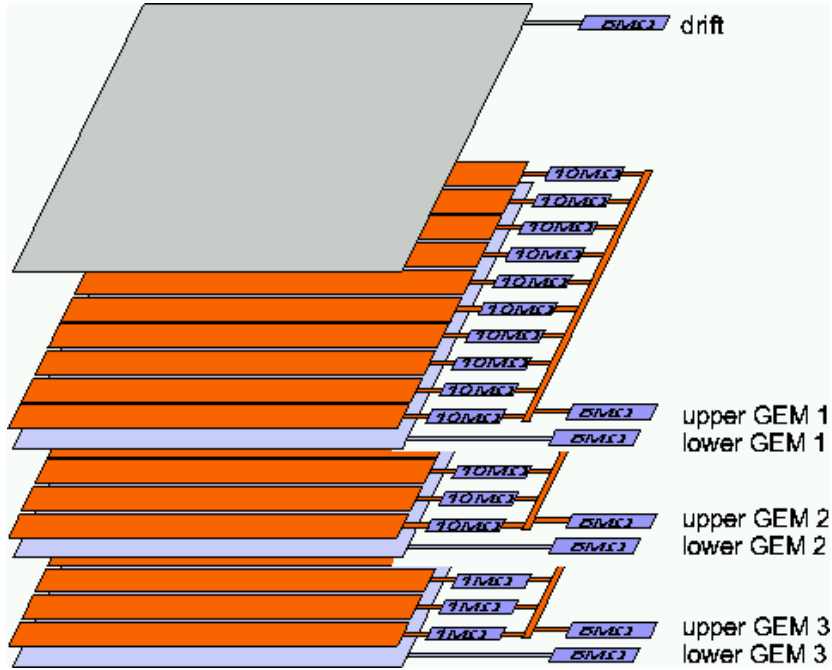}
\caption{HV-schema, each of the seven needed voltages is delivered by 
individually controllable power supply channels.}
\label{hv-sceme}
\end{figure}

To allow for maximal flexibility in choosing the electrical fields that are
needed to run the detector all high voltages are delivered separatly by an
individually controllable power supply channel.  The high voltage schema is
shown in figure \ref{hv-sceme}.  The currents are monitored for each high
voltage channel. A discharge in a GEM can be indentified by observing the
reloading current of this GEM segment.

Of the two detectors built, one has the segmented GEM on the upper
side as in figure \ref{hv-sceme}, in the other detector the segmentation
is on the side towards the readout board. No significant
difference (e.g. in sparking probability) has been observed.

\subsection{Two dimensional readout board}

\begin{figure}
\begin{minipage}[t]{.47\textwidth}
\begin{center}
  \includegraphics[width=1.0\textwidth]{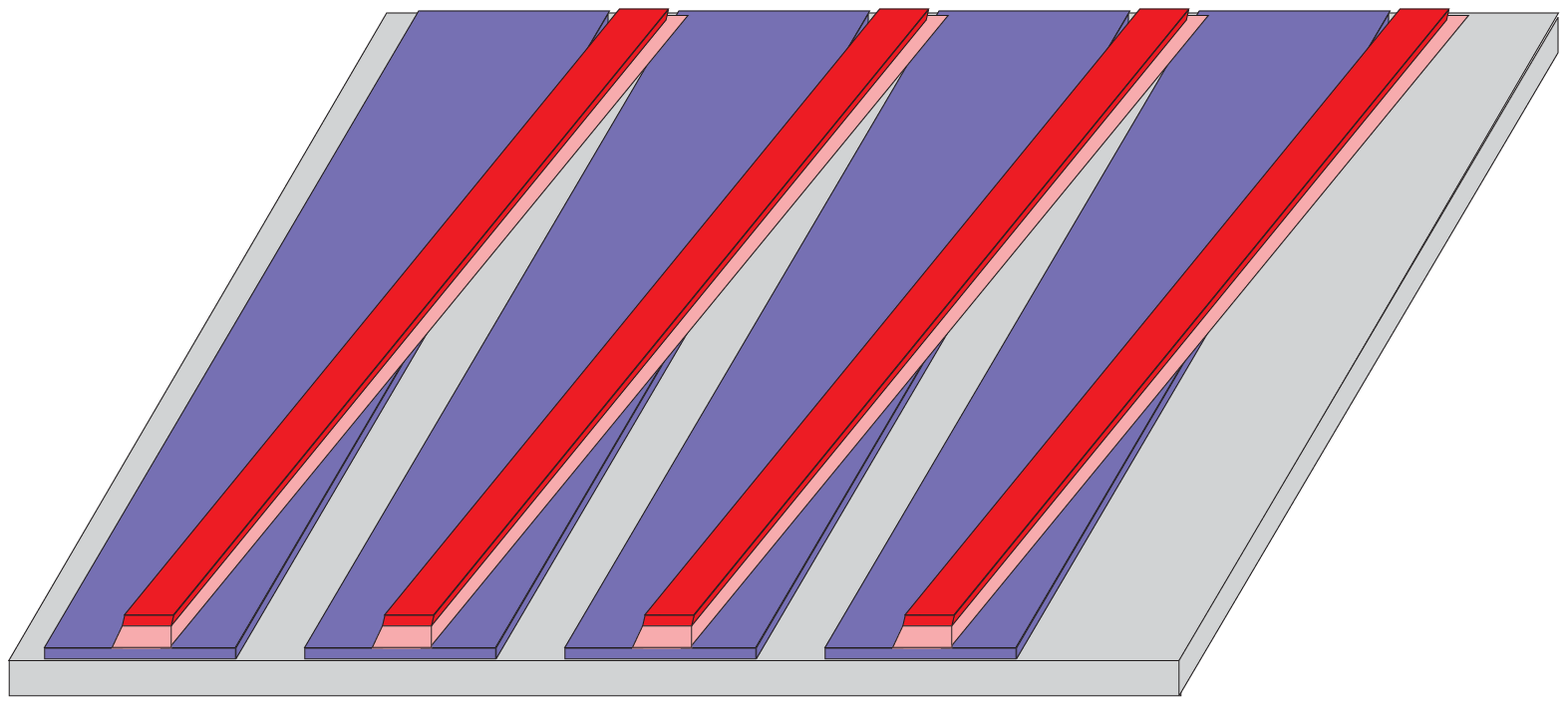} \end{center}
\caption{Schematic
    drawing of the readout board. Pitch 300 $\mu$m, upper strip width 60
    $\mu$m, lower strip width 250 $\mu$m.} \label{readout-pcb}
\end{minipage}\hfill
\begin{minipage}[t]{.47\textwidth}
\begin{center}
\end{center}
\caption{An optical microscope picture of the readout board.}
\label{strips-microscope}
\end{minipage}
\end{figure}

The separation of the amplifying stage from the readout stage in a triple GEM
detector allows to use any readout pattern that is appropriate for the
application.  Here a two dimensional readout printed circuit board was chosen with a stereo
angle of $5^{\circ}$ between two sets of parallel strips.  The rotated strips
on the top are separated through a 50 $\mu$m Kapton layer from the lower
strips (see figure~\ref{readout-pcb} and figure~\ref{strips-microscope}).  The
lower strips have to be wider than the upper ones in order to collect in
average an equal amount of charge on both layers (charge sharing).  We have
used 60 $\mu$m and 250 $\mu$m wide strips with a pitch of 300 $\mu$m for each
plane. Although this pitch was chosen for practical reasons to connect the printed circuit board
to a HELIX front-end chip, it turns out to be reasonable choice for such
a detector (see the discussion of the results below). The length of the strips
is 30 cm.

This readout board was manufactured from copper cladded Kapton foil with the
same etching technology as the GEM foils in the CERN workshop. It is glued to
a 100 $\mu$m thick G10 sheet to insure a reasonable flatness. Below the
readout board a 4 mm thick Rohacell plate is mounted covered with a grounded 
aluminum foil, which acts as an electrical shield.

\subsection{Readout electronics}

The readout strips are connected to the HELIX 128 readout chip \cite{helix},
originally developped
%\marginremark{stimmt die HELIX Version?}  
for the HERA-B experiment.  Each layer of a detector is equipped with 4 chips
(= 512 read out channels). Two thirds of the active area of the detector can
be read out this way.

The connection (see figure \ref{helix-seite}) is made through a short (3 cm
length) Kapton stripline bonded with Z-bond glue\footnote{done by Dr. U. Werthenbach
  University of Siegen} and a fan-in, which fits the pitch of the HELIX input
pads of 41.4 $\rm\mu m$.  This fan-in is made of thin film
ceramics\footnote{manufactured by Siegert TFT, Hermsdorf, Th\"uringen} and
includes a serial overcurrent protection resistor of 600~$\Omega$ to avoid
damage of the preamplifier input in case of chamber sparks or shorts
(optimised for the HERA-B MSGC operation).

The signal path is closed through the capacity between the lowest GEM and the
readout board of about 500 pF.

\begin{figure}[!b]
\begin{center}
\caption{HELIX boards connected to
    readout PCB via a Kapton fanout.} \label{helix-seite}
\end{center}
\end{figure}

The capacity of the ceramic fanin was measured to be 2 pF. The total capacity
of the fully mounted system of one strip to all its neighbour strips and the
full electrical environment measured by coupling a square wave pulse to the
readout board (still connected to the HELIX) through a 1 k\Ohm serial
resistor.  From the observed rise time of this pulse a capacity of 86 pF was
determined.

\begin{figure}
\centering
\includegraphics[height=7.5cm]{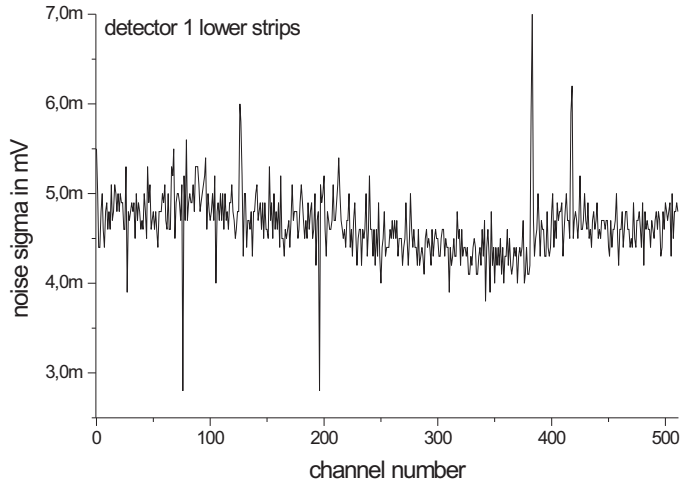}
\caption{Noise distribution for the readout channels in one readout
  layer. Channels with lower noise indicate broken readout strips, bad
  connection to the electronics or damaged readout channels. A higher noise
  can indicate a short between readout strips.  } \label{noise-sigma}
\end{figure}

By coupling a well defined charge to a readout strip the HELIX gain was
measured to be 880 e$^-$/mV in this configuration. The average noise level
(Figure \ref{noise-sigma}) of 4.8 mV corresponds therefore to 4200 electrons.
From this information and the HELIX data sheet a total capacity of the readout
board of $\approx$ 100 pF was derived, which is consistent with the value
above. In addition an electrical simulation of the setup gave similar figures
on the capacity \cite{lev}.

The HELIX comes in its analog characteristics very close to the chip, that will
be used later in LHCb.  The shape of the preamplifier output of the HELIX
depends on the capacity of the detector connected to the input. For our
detector it has a FWHM of about 90 ns. The amplitudes of these signals are
sampled by a 10 MHz clock and are stored in the analog pipeline, that is
included in the HELIX chip. The HELIX was run in standard conditions \cite{helix}.

The 10 MHz clock was in general operated independantly of the particles
traversing the chambers. Therefore the sampling time was not optimally set to
the pulse maximum, but randomly distributed over 100 ns.
The data used to measure the pulse height distributions were however taken
with an additional trigger condition requiring the signal to be within 
25~ns of the active clock edge.

After a trigger occurs each HELIX chip sends the analog value from all 128
channels multiplexed through one readout line, which were digitised by an
oscilloscope connected to a PC via GPIB bus and controlled by LabVIEW. The
communication between the oscilloscope and the PC limited the readout speed to
0.5 Hz.  For cosmics data taking the scope was replaced by VME ADC
boards.

\begin{figure}[ht!]
\centering
\includegraphics[width=0.49\textwidth,height=5.4cm]{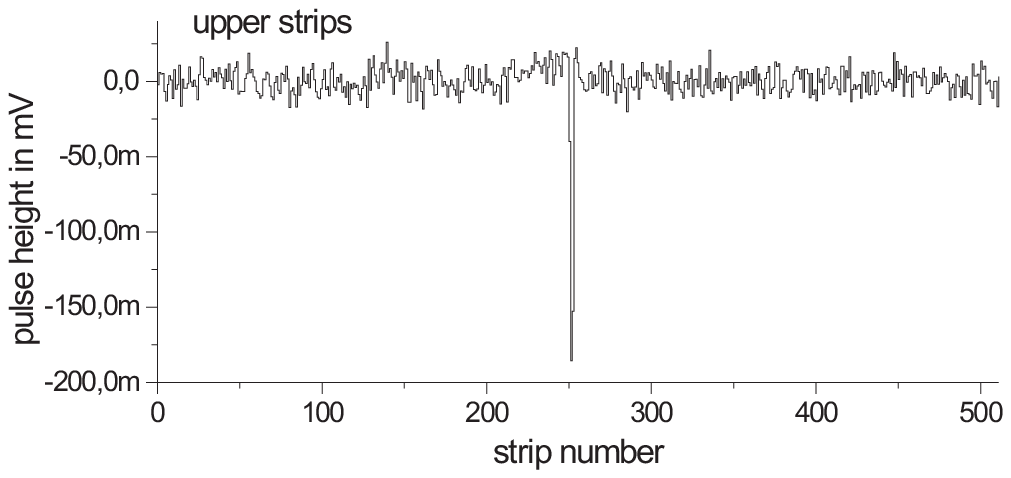}
\includegraphics[width=0.49\textwidth,height=5.4cm]{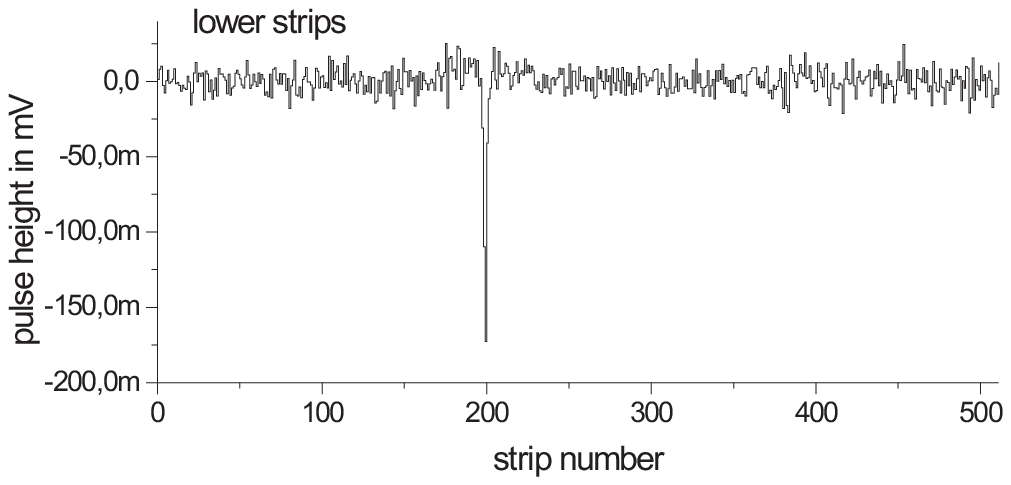}
\caption{Corrected Helix data from a minimum ionising particle. The signals
  of the two readout coordinates as a function of the channel position are
  shown.}
\label{up-low}
\end{figure}

The pedestal for each readout channel is subtracted offline and each signal
is corrected for baseline variations (common mode noise) by using a filter to
remove large period components from the data.  In figure~\ref{up-low} the
corrected data from a minimum ionising particle crossing the detector is
shown. For identifying the signals a threshold of 3.5 sigma of the
noise distribution is applied.

\subsection{Signal shape}
In order to study the shape of the chamber signal a fast amplifier
VV61\footnote{R. Rusnyak Physikalisches Institut der Universit\"at
  Heidelberg} with a rise time of 0.5 ns was connected to the detector.

\begin{figure}
\v{0.5}
\begin{minipage}[b]{.5\linewidth}
\begin{center}
\epsfig{file=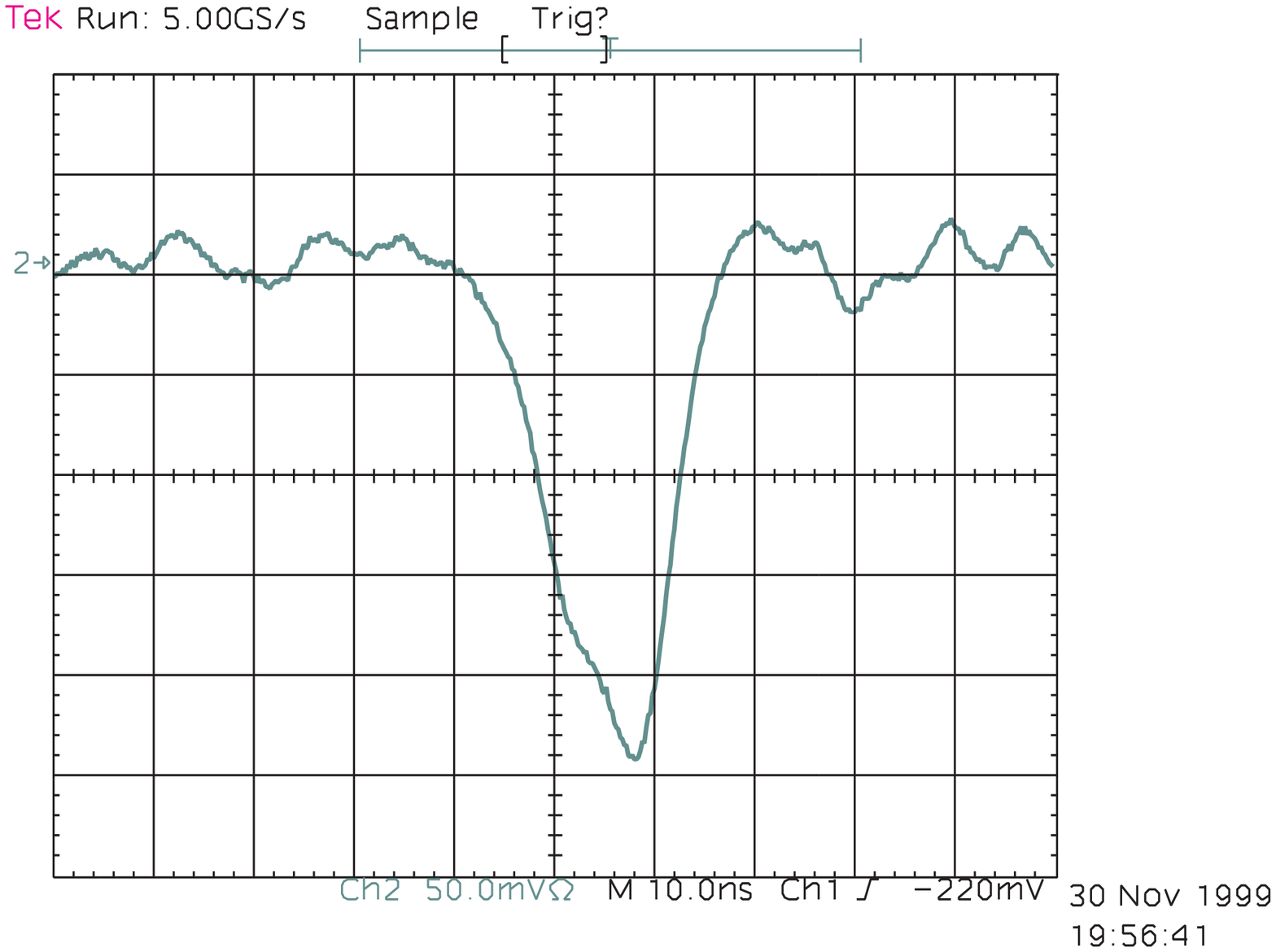,width=0.8\linewidth}
\end{center}
\end{minipage}\hfill
\begin{minipage}[b]{.5\linewidth}
\begin{center}
\epsfig{file=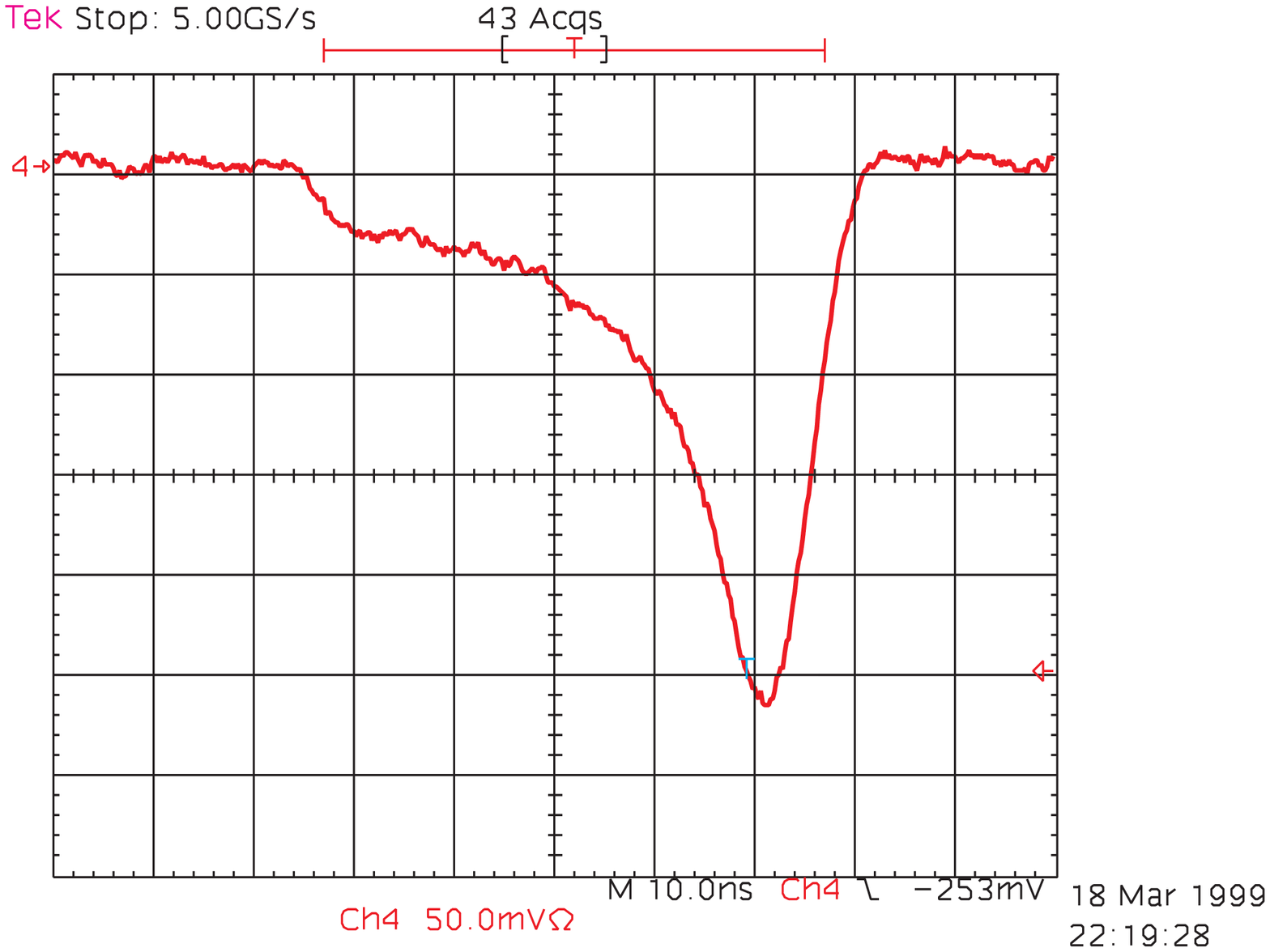,width=0.8\linewidth}
\end{center}
\end{minipage}
\caption{Left: signal shape for the new detector with 1mm transfer gap 
  (8 strips connected together).  Right: signal shape of the old detector with
  3 mm transfer gap (5 strips connected together)}\label{vv61}
\end{figure}

The signal from a Fe$^{55}$ source is shown in figure~\ref{vv61} on the left.
For comparison the signal of our previous prototype detector \cite{triplegem1}
with a transfer gap of 3 mm is shown on the right in the same figure.  The
pulse shape can be understood by interpreting the signal as the induced
current due to the moving charged cluster in the gap between the lowest GEM
foil and the readout board: the drift velocity in Ar:CO$_2$ (70:30) at 5 kV/cm
is $\approx$7.5 cm/$\mu$s, corresponding to a drift time for the 3 mm gap of
40 ns. Although the total current should be constant, we can observe a rising
signal due to the effect that the strips close to the center of the cloud see
an increasing fraction of the total current, as the charged cloud moves closer
to the readout board surface. This rise time is close to the expected 40 ns.

For the new detector the rising of the signal is barely visible, since the
drift time in the 1 mm gap is expected to be about 13 ns only.

\section{Operation and performance measurements}

Two identical complete detectors were built and mounted back to back, such
that the stereo angles are +5$^{\circ}$ and -5$^{\circ}$ compared to the
vertical strips.  The distance between the centers of the primary ionisation
gaps is 17.9 mm (see figure \ref{setup} right) allowing a crude crossing 
angle determination of the observed tracks.

The detectors were operated both at PSI beams and in a cosmic ray setup. 

\subsection{High rate particle beam experiment}
\begin{figure}[ht!]
\centering
\mbox{\ }\hfill
\includegraphics[width=0.46\textwidth]{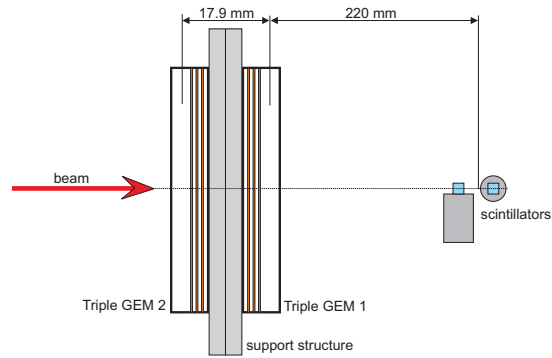}
\caption{Left: setup at PSI in the $\pi$M1 area.
  Right: schematic drawing of the chamber arrangement and the scintillators.}
\label{setup}
\end{figure}

At PSI four other groups of the LHCb Inner Tracker collaboration tested their
detectors simultanously in December 1999.  Picture~\ref{setup} (left) shows
the triple GEM detectors mounted between the prototypes of the other groups.

Beam intensities and distributions were measured by a pair of 5$\times$5
mm$^{2}$ scintillators mounted on a remotely movable table. The width of the
beam was typically $\approx$6 cm. For operation stability tests a 350 MeV/c
$\pi^+$ beam was used. The maximum achievable total rate was 60~MHz with a peak
rate of 1$\times$10$^4$ Hz/mm$^2$. For most of the data taking a 215 MeV/c
$\pi^-$ beam with low intensity (60 Hz/mm$^2$) was used.

\subsection{Cosmic ray experiment}

For a high statistics study of the performance as a function of the
track crossing angle a cosmic ray setup was used. The chamber were
mounted horizontally. A pair of 10$\times$10 cm$^2$ scintillators were
mounted below and above the detectors. To be able to study large
clusters in detail the chambers were operated at a higher
gain ($V_{GEM}=390$ V) than usually.

\subsection{Electrical field configuration}

As an optimal HV setting the following values were exposed: drift field 3.0
kV/cm; transfer fields (between the GEM foils) 2.2~kV/cm; collection field 5.0~kV/cm. 
The most critical fields with respect to sparking are the two transfer
fields. An increasing of these fields from 2.2 kV/cm to 3.0 kV/cm lead to an
enhancement of the sparking probability by a factor of 10.

At the beginning the detector was operated with a drift field of 2.0 kV/cm.
Later we discovered that an increase to 3.0 kV/cm causes a gain reduction of
15 \%. This is qualitatively explained by a better transparency for lower
drift fields: in the optimal case all field lines from the drift cathode are
focused and compressed through the holes terminating on the readout board.
According to simulations in the symmetric case a typical transparency for the
primary electrons of 88\% \cite{cwetanski} can be reached in our
configuration. By increasing the drift field some field lines terminate on the
GEM foil and primary electrons are therefore lost.

It was observed, that the sparking probability is also 
reduced by about a factor of three with higher drift field corresponding
to the lower gain of the chamber (see section on gain and spark probability).
In contrast to that, the collection field between the lower side of the 
last GEM foil does not seem to influence the sparking rate.
 
%This may be explained by the effect that more 
%of the back drifting ions are accelerated directly to the drift electrode 
%instead of being catched by the GEM.  Ions that are catched by the upper 
%GEM side kick out electrons that cause further
%ionisation (spontaneous field emission).  
%This feed back effect may result in
%a GEM spark.
%\marginremark{Also das glaube ich ueberhaupt nicht, wir reden doch die ganze
%  Zeit von streamer discharge, oder?}

After the amplification process in a GEM hole a major part of the electrons is
lost to the lower GEM side.  If the field on the lower side is increased, more
electrons can be collected to the readout board ({\it collection efficiency}).
On the other hand a high collection field may cause sparks to propagate to the
readout board and may damage it.  Because of the robustness of the readout
board and the well proven high voltage protection of the HELIX readout
electronics, we decided to increase the collection field to 5.0 kV/cm.

According to simulations the collection efficiency in this configuration for
the upper and middle GEM is 42 \%, for the lowest GEM 62 \%. Taking into
account in addition the primary electron transparency and the loss due to
attachment in the gas, the total fraction of all primary and secondary
electrons collected on the readout board is only 7.5 \% \cite{cwetanski}.

\begin{figure}[htb]
\centering
\includegraphics{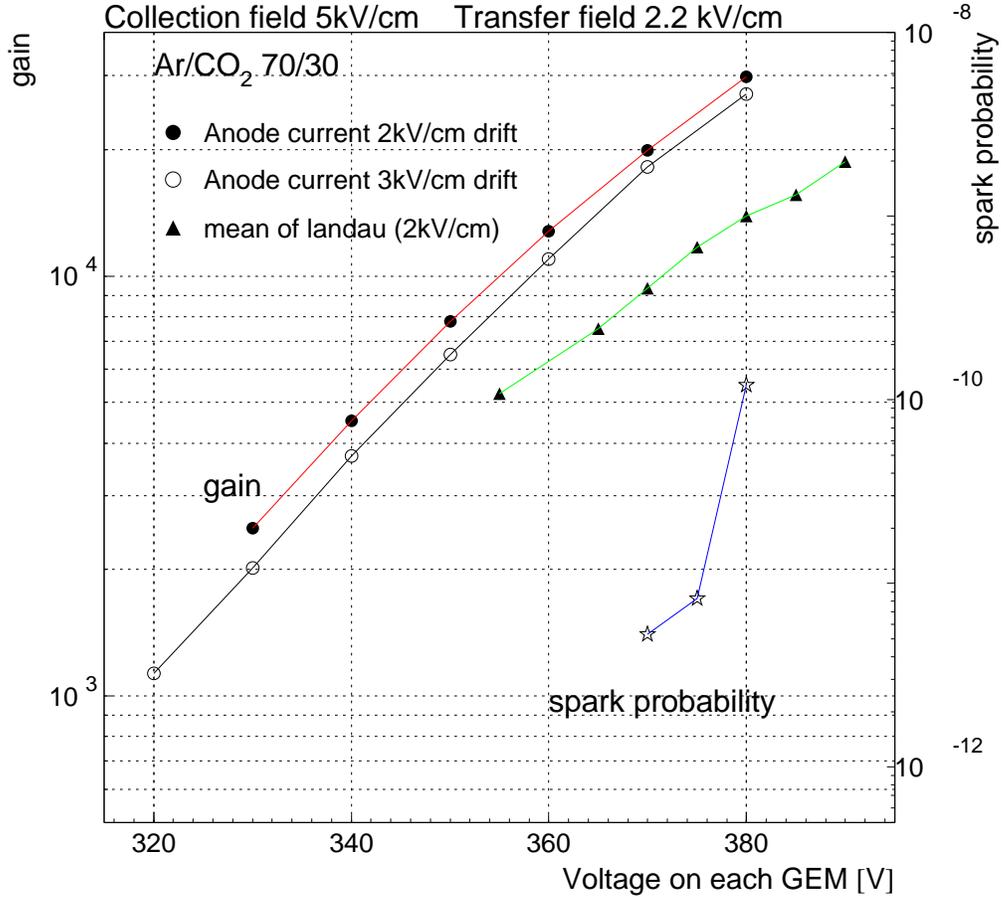}
\caption{Detector gain, determined with different methods, as a function
of GEM voltages, where all three GEM are set to the same voltage.
Also shown is the spark probability per incoming particle}
\label{fig:gain}
\end{figure}

\subsection{Detector signal gain and spark probability}

The detector is operated with Argon and CO$_2$ gases with a mixing ratio of
70:30.  Although CO$_2$ alone is not a very stable quencher gas, the simpler
infrastructure with respect to safety and the chemical ageing stability of this
mixture outweighs a possible improvement in discharge performance of more
complex quencher gases. The primary ionisation gap has a width of 3~mm,
where
in the Ar:CO$_2$ 70:30 gas mixture a vertically incidenting minimum ionising
particle produces in average 21 primary electrons in about 8 clusters
\cite{cwetanski}.

The detector gain (see figure~\ref{fig:gain}) was measured in various
different ways. 

The total incident particle rate was calculated by measuring the coincidence
rate of the two scintillators moving them along $x$ and $y$ across the beam
cross section.  The anode current was determined by measuring the current in
the ground lines of the preamplifiers and it was verified, that this is the
same value as the sum of all upper and lower GEM currents and the drift
current taking the proper sign into account.  From this current and the
assumption of 21 primary electrons per minimum ionising particle the detector
gain was calculated.

Due to a possible missalignment of the scintillators against each other
an overestimation of the rate of 20 \% is possible. The exact
shape of the beam varies with details in the beam line settings giving
rise to a further systematic uncertainty of about 20 \%. 

The gain was also determined from a Landau distribution fitted to the pulse
height spectra (see section on results) and the known gain of the electronics
chain taking into account the charge sharing between the two readout
coordinates. 

Furthermore signals of an Fe$^{55}$ $\gamma$--ray source were observed in the
chamber resulting in gain values between the two previously mentioned. In
summary, considering the various uncertainties we believe, that the
correct signal gain is located somewhere between the two curves in
figure~\ref{fig:gain}.

When turning on the detector the gain observed is initially lower and slowly
increases with time during irradiation with a source. After a few minutes the
signal size becomes stable at about double of the initial value. This is
interpreted as charge up of the Kapton surface in the GEM holes and has been
seen by many other groups as well \cite{malte}.

The spark probability per particle was measured at 350 MeV/c beam energy (see
figure~\ref{fig:gain}).  The maximal beam intensity was 60 MHz.  For example
45 sparks were observed in both detectors in 12.5 hours at a GEM voltage of
375 V, resulting in a spark probability of 8.3$\times$10$^{-12}$.

\section{Results}

\subsection{Analysis procedures}

After the detector signals were digitised and stored on disk the analysis
proceeds in an offline procedure with the following steps:

\begin{enumerate}
\item Baseline correction and subtraction as described in the section of
  readout electronics.
\item Applying a threshold of 3.5 times the average noise. Each channel is
  treated seperately. Signals exceeding this threshold are identified as a
  cluster.
\item The center of gravity and both the root mean square of the width and the
  number of channels above threshold of these clusters are determined.  For
  the PSI data in addition a Gaussian fit to each signal gives the width of the
  cluster in terms of $\sigma$ in units of strips.
\item In cases, where there is exactly one cluster in each plane, the $x$ and
  $y$ coordinates are calculated from the center of gravities.  This works
  only for the low rate data at PSI and for cosmics.
\end{enumerate}

\begin{figure}
\centering
\includegraphics[width=0.60\textwidth]{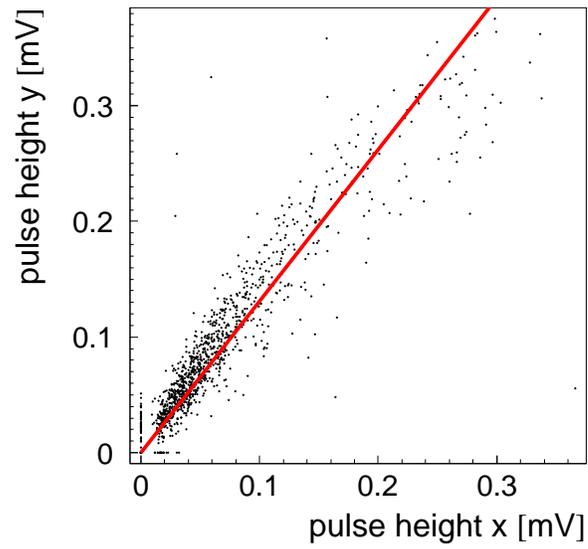}
\caption{Correlation between pulse heights of the 
lower (x) and upper (y) strips.}
\label{fig:cha}
\end{figure}

\subsection{PSI data}

\subsubsection{Pulse height and efficiency}

Figure~\ref{fig:cha} shows the correlation between the pulse heights of the
two readout coordinates as determined from low rate PSI data. In average the
signals on the upper strips are $\approx$1.25 times larger than those of the
lower ones.

\begin{figure}
\centering
\includegraphics[width=0.70\textwidth]{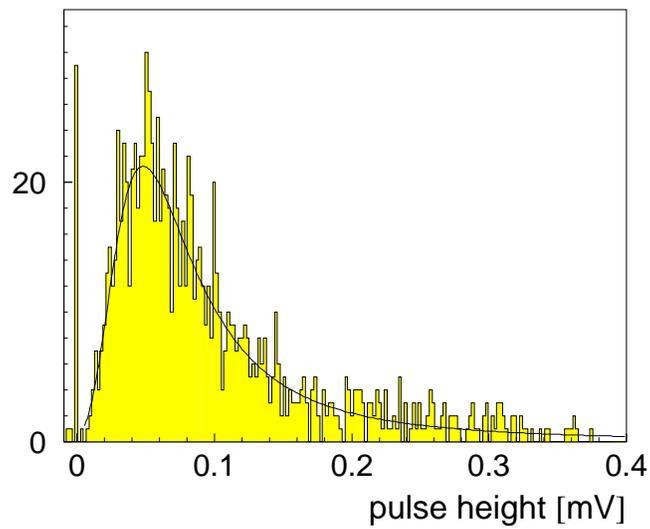}
\caption{Pulse height distribution measured at a GEM voltage of 380 V.
Signals below the threshold are filled into the bin at zero.}
\label{fig:lan}
\end{figure}

Figure \ref{fig:lan} demonstrates, that the pulse height distribution can be
well fitted by a Landau type function reflecting the pulse height variations
due to primary ionisation statistics. These distributions were used to
determine the gain (see section on chamber gain).

\begin{figure}
\centering
\includegraphics[width=0.70\textwidth]{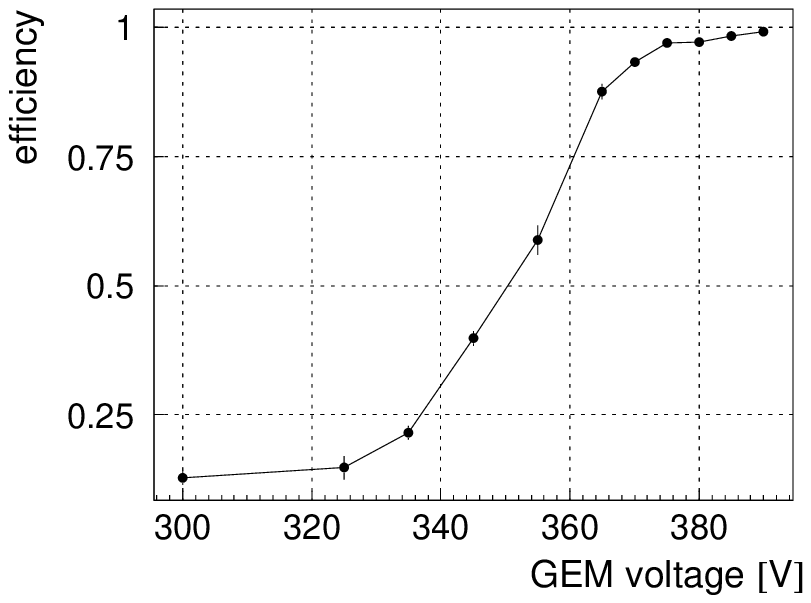}
\caption{Efficiency as a function of the GEM voltage.}
\label{fig:effi}
\end{figure}

Counting the number of non empty events for each scintillator trigger
and variing the GEM voltages allows to determine an efficiency plateau
for each of the four available planes (two chambers with two readout
coordinates each), one of those is displayed in Figure \ref{fig:effi}.  

\subsubsection{Cluster size}

\begin{figure}
\centering
\includegraphics[width=0.70\textwidth]{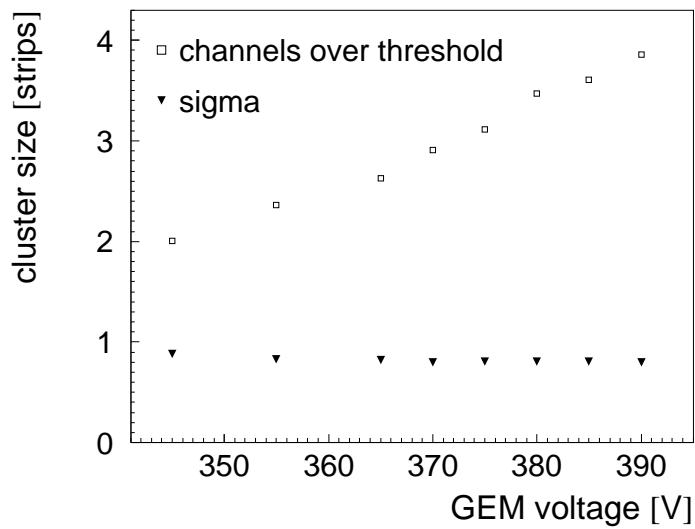}
\caption{Cluster width as a function of detector gain, determined from 
the number of channels above the 3.5 sigma noise threshold (open squares) 
and from the width of a Gaussian fit to the pulse shape (triangles).}
\label{fig:clustersize}
\end{figure}

The width of the signal was studied as a function of
the GEM gain. In figure \ref{fig:clustersize} two different methods
of determining the cluster size are compared. An often used definition
just uses the number of channels in the cluster, which have a signal
larger than the noise threshold. This quantity clearly depends on the
pulse height and as expected the average values rise with the gas gain. 

A more physical definition of the cluster size uses the $\sigma$ of a Gaussian
fit to each pulse. Figure \ref{fig:clustersize} shows, that this cluster size
does in fact not depend on the gain of the chamber.  The value of 0.8 strip
pitches (240 $\mu$m) is consistent with an estimation, which is based only on
the transvere diffusion ($D_t \approx 300 \;\mu\mbox{m}/\sqrt{\mbox{cm}} $) of the
charged cloud along an average drift length of 4.5 mm.

\begin{figure}
\centering
\includegraphics[width=0.49\textwidth]{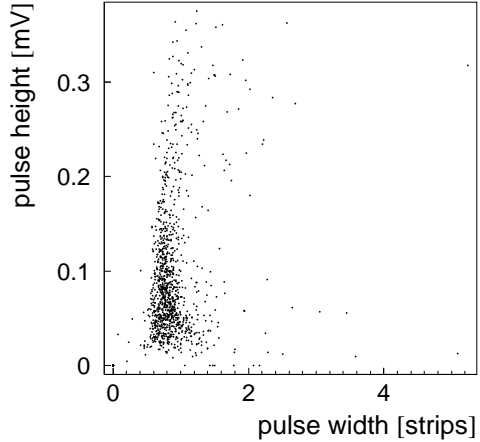}
\caption{
Correlation between pulse width and height.
To determine the width of the signal a Gaussian fit was used.}
\label{fig:width}
\end{figure}

In addition the cluster size (using the $\sigma$ from the Gaussian fit) is
correlated with the pulse height.  A substantial electronic crosstalk between
neighbouring readout strips would manifest itself in an increasing width for
larger pulses.  Figure~\ref{fig:width} indicates, that there seems to be no
problem with electronic crosstalk.

\begin{figure}
\centering
\includegraphics[width=0.49\textwidth]{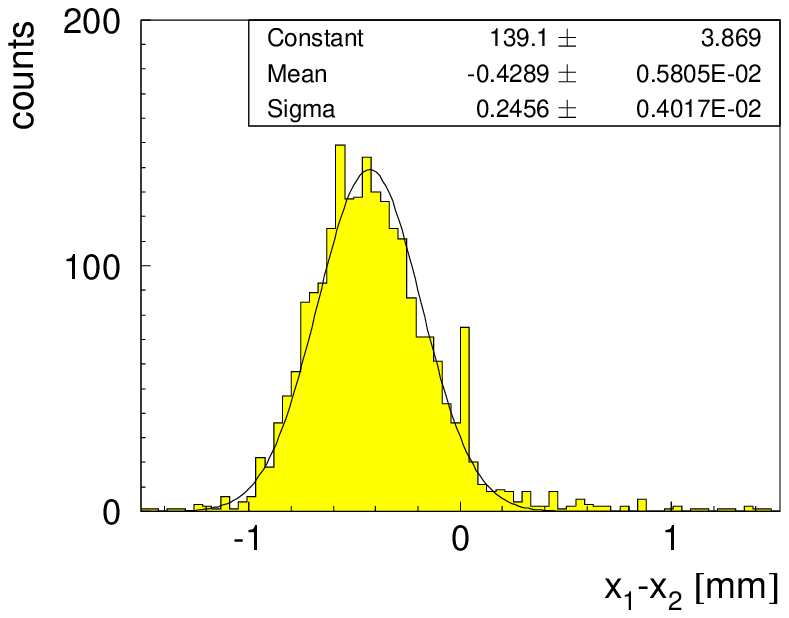}
\includegraphics[width=0.49\textwidth]{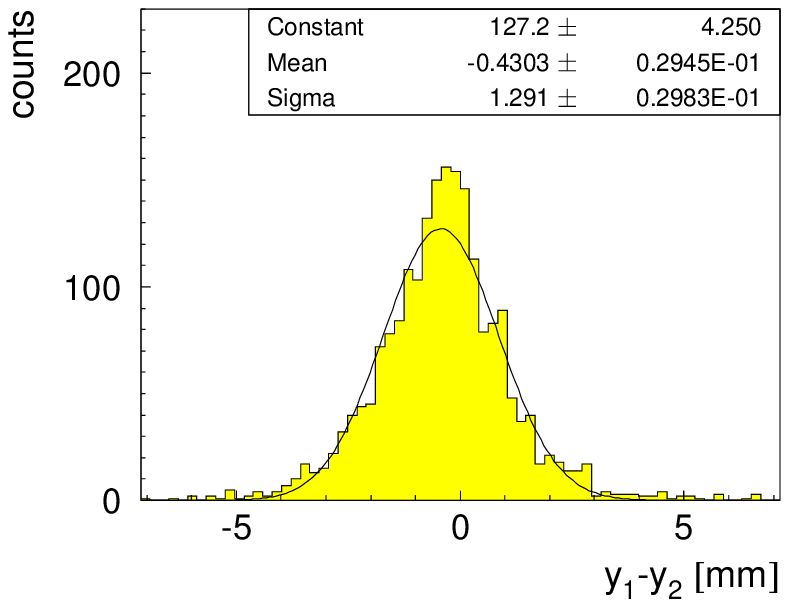}
\caption{Difference between the x-coordinate (left) and
the y-coordinate (right) of the two detectors, measured at 370 V 
GEM voltage}
\label{fig:res}
\end{figure}

\subsubsection{Resolution}

Although 215 MeV/c $\pi$'s give large multiple scattering and are therefore
not suitable to do a resolution measurement, the differences between the
coordinates measured by the two detectors are plotted as a cross check in
figure~\ref{fig:res}. It shows that the alignment of the two detectors is well
understood in both coordinates. A Gaussian fit gives $\sigma_x \approx$ 0.25
mm and $\sigma_y \approx$ 1.3 mm.  Multiple scattering in the first chamber
contributes about 0.1 mm to these values, while the angular divergence of the
beam is estimated to be below 20 mrad, an upper limit of its contribution is
therefore 0.36 mm indicating, that $\sigma_x$ is dominated by this beam
divergence. For a real spatial resolution measurement a better
defined beam of higher energy is obviously needed.

\subsection{Cosmic ray data}

\begin{figure}
\centering
\includegraphics[width=0.49\textwidth]{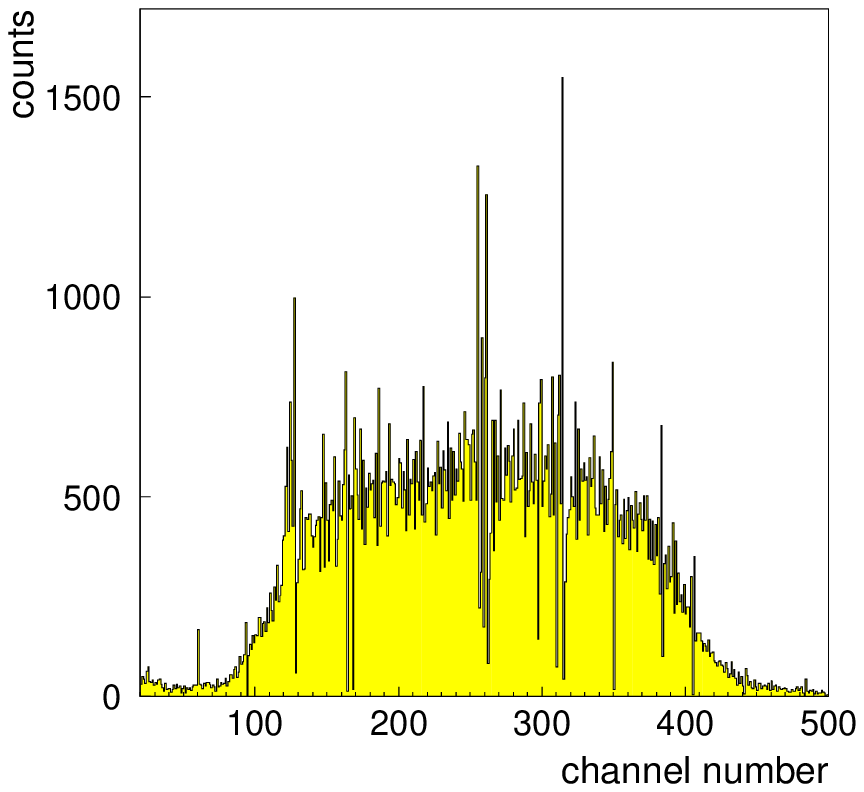}
\includegraphics[width=0.49\textwidth]{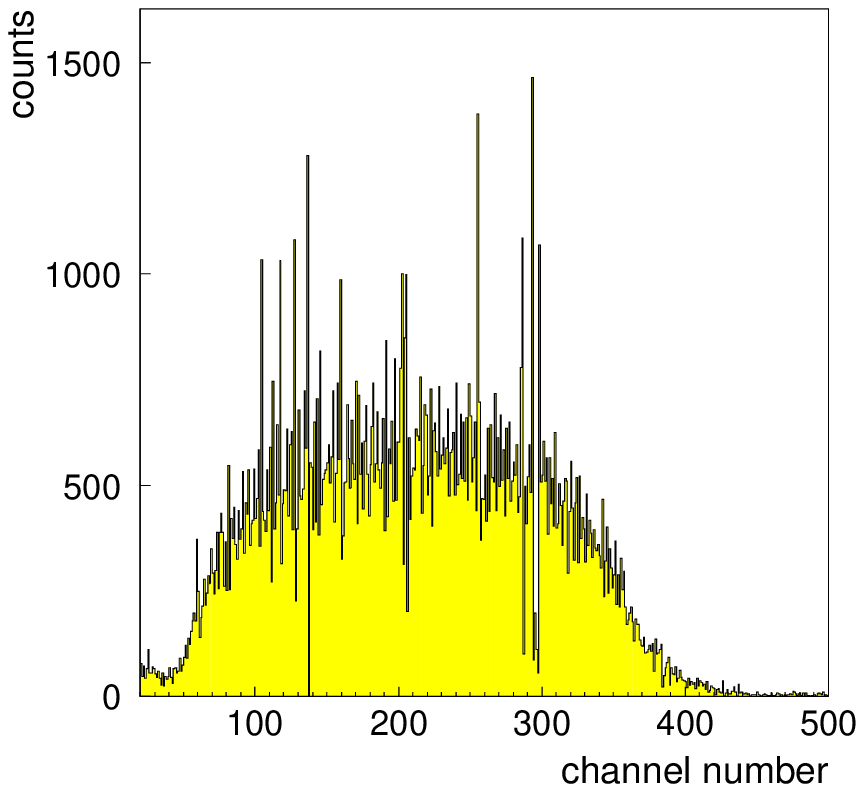}
\caption{Channel map of lower (left) and upper strips (right).}
\label{fig:wire}
\end{figure}

With the identical detectors also cosmic data was taken.  In
figure~\ref{fig:wire} the channel map for the lower and upper strips is shown.
Figure~\ref{fig:pos} shows the reconstructed hits of all recorded cosmic data.
To cover a larger area of the detector the scintillators were moved gradually
in $y$-direction.  The distinct region with no hits corresponds to a
disconnected GEM segment. Broken and hot channels (masked off) are clearly
recognised.

\begin{figure}
\centering
\caption{Hit map of recorded cosmic data.
Broken strips are clearly visible.}
\label{fig:pos}
\end{figure}

\subsubsection{Signal homogenity}

\begin{figure}
\centering
\includegraphics{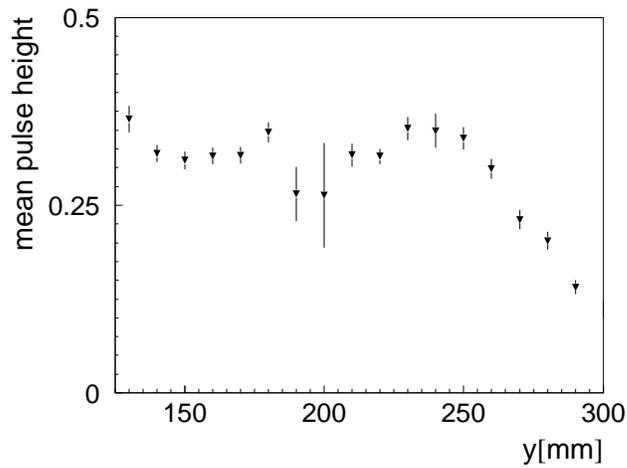}
\caption{Signal gain homogenity. The average pulse height is shown
  as a function of the position.  At $y=300$ mm the frame of the chamber is
  located. }
\label{homo}
\end{figure}

The homogenous irradiation of the chamber with high statistics allows to
observe variations in the signal size in different regions of the chamber. In
general the detectors behave rather homogenous, however closer to the borders
of the active region the signal size starts to drop down to about a half of
its value compared to the center of the detector (see figure \ref{homo}).
Similar reductions in signal size have been observed on all chamber
boundaries.

There is a suspicion, that the copper and the Kapton GEM holes might be badly 
aligned to each other at the border of large surfaces due to stability
problems of the film masks \cite{alignment}.

\subsubsection{Track inclination}

\begin{figure}
\centering
\includegraphics[width=0.47\textwidth]{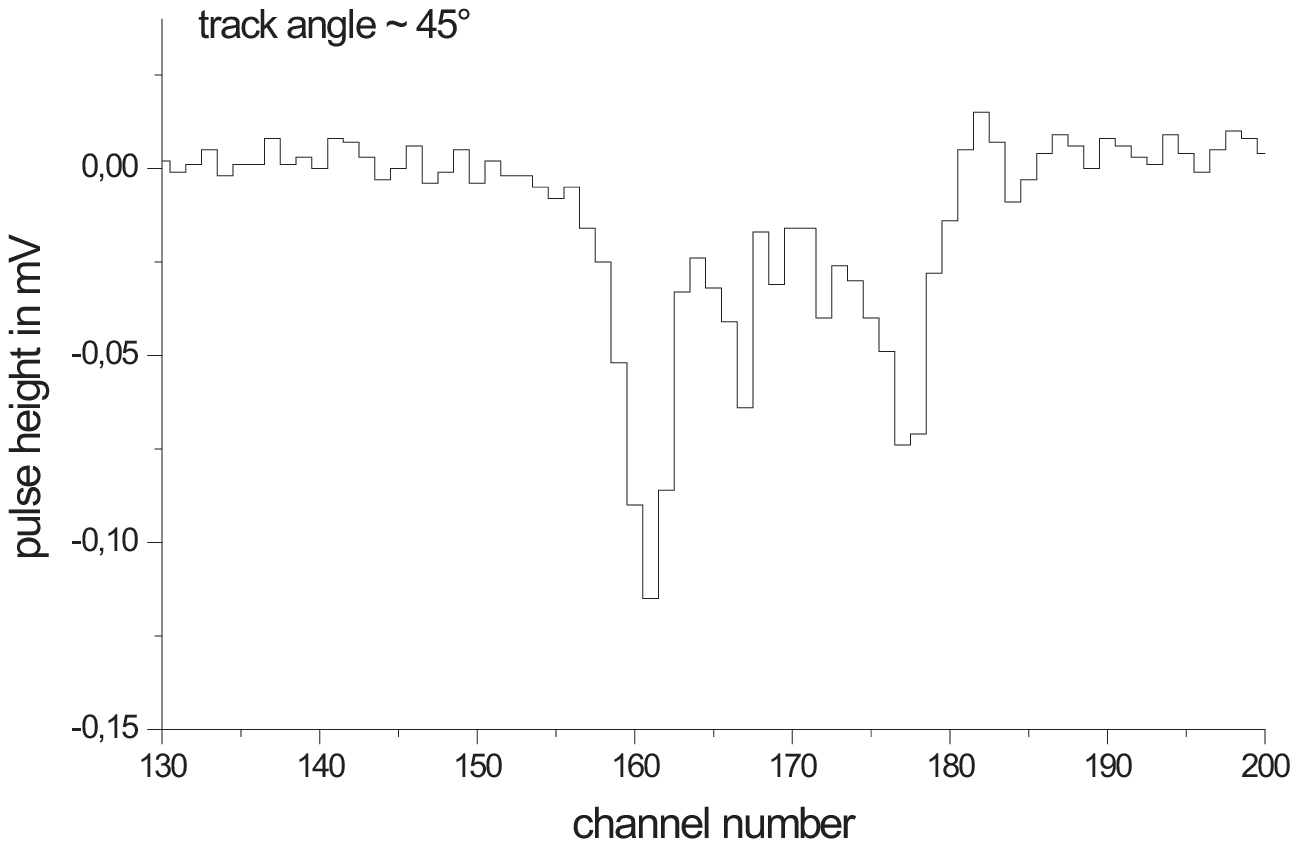}
\hfill
\includegraphics[width=0.47\textwidth]{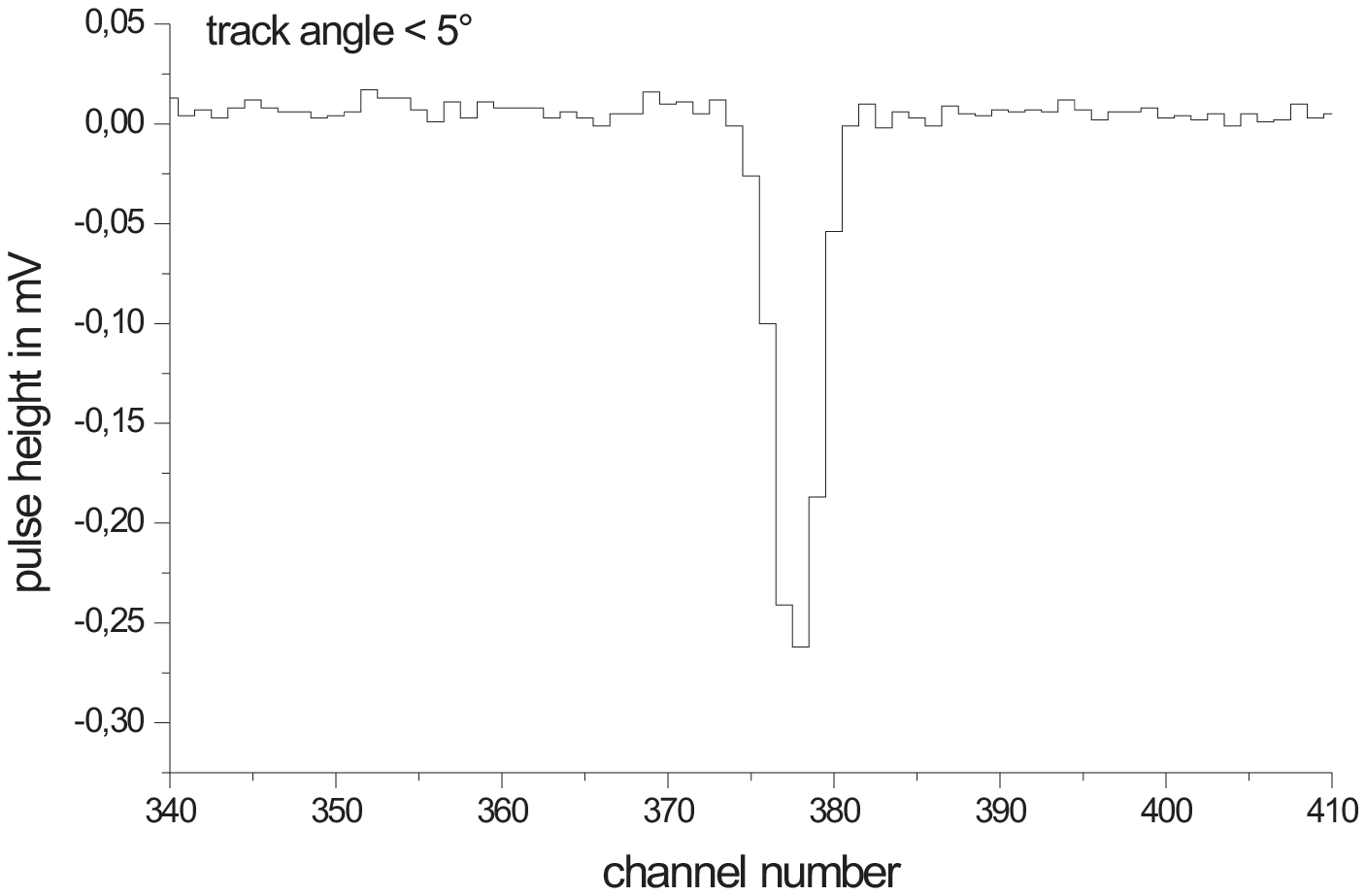}
\caption{The signal size as a function of strip position for a single
  cosmic ray event with an inclination angle of about 45$^\circ$ (left).  For
  comparison the figure on the right shows a signal from a vertically
  incidenting track.}
\label{fig:eventlowangle}
\end{figure}

\begin{figure}
\centering
\includegraphics[width=0.70\textwidth]{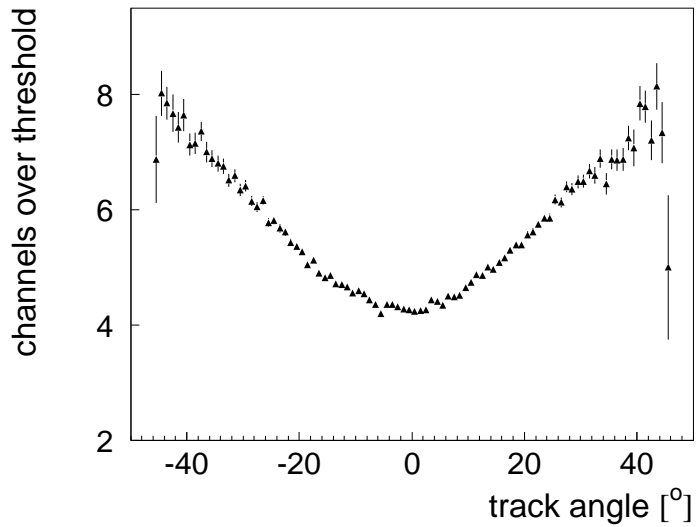}
\caption{The clustersize of the cosmic ray data as a function of
the inclination angle.}
\label{fig:angle}
\end{figure}

\begin{figure}
\centering
\includegraphics{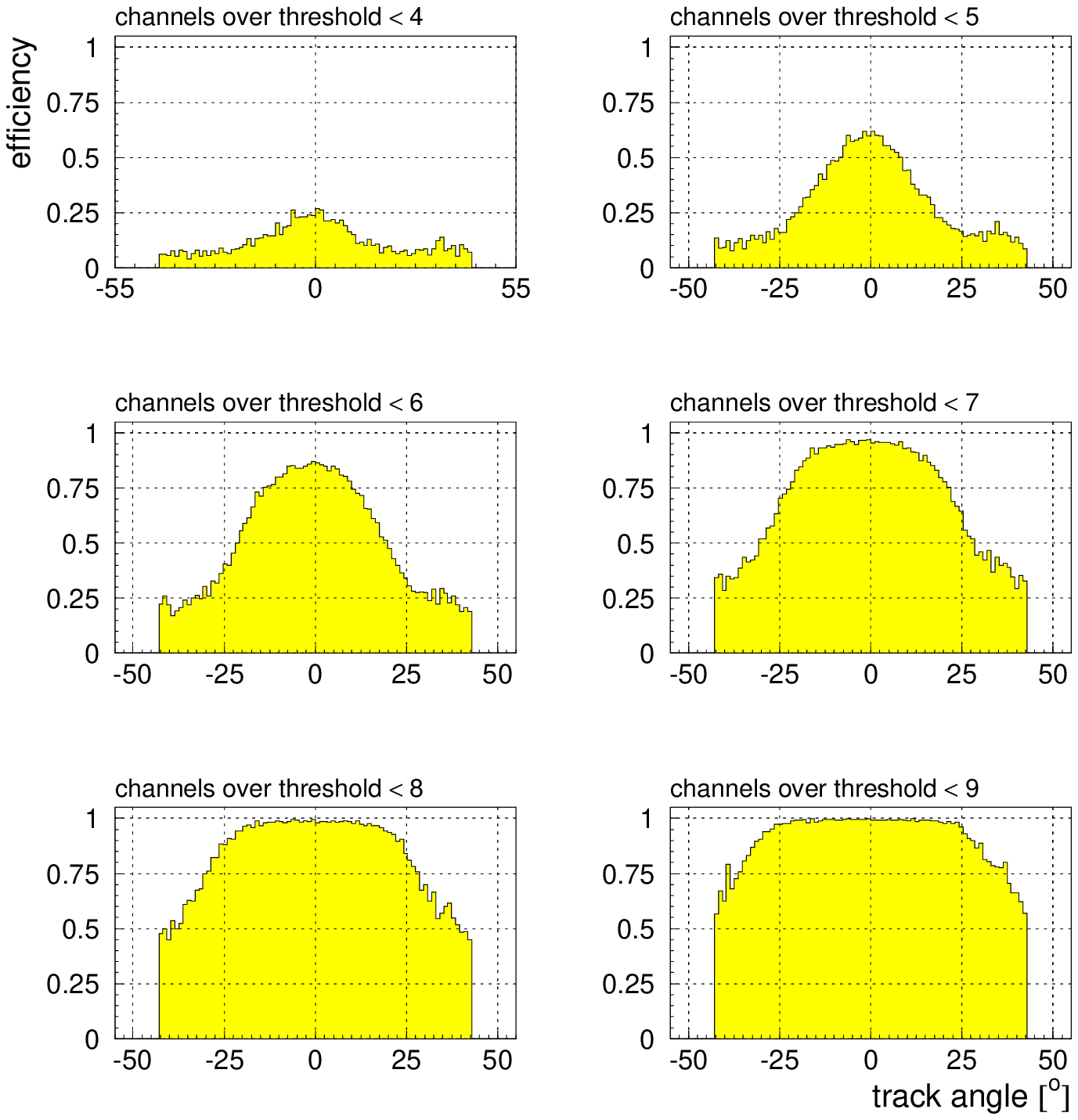}
\caption{Detector efficiency as a function of track inclination angle for
different threshold cuts in ADC units}
\label{fig:eff2}
\end{figure}

In the LHCb application the distribution of the inclination angle
of the dominating background particles (mainly electrons and positrons)
is very broad. It is therefore interesting, how the signals depend
on the polar angle of the incident particle.

The cosmic ray sample is well suitable to study this question: from the 
position information of the two detectors the 
inclination angle of the incoming cosmic particle was reconstructed, and
the events were studied as a function of this angle.

Figure \ref{fig:eventlowangle} shows the signal of a cosmic ray event observed
under an angle of about 45$^\circ$ compared with a normal, vertical
incidenting track. The cluster is wider as expected since the tracks crosses
several drift cells. The signal height varies according to the primary
ionisation statistics. Due to the smaller number of clusters for each cell
this variation is much bigger than for a normal signal. This leads to the
difficulty that from the signal alone it is impossible to decide whether
the event originates from a single, inclined track or from several vertical
particles. In a high rate application additional tracking information from
other devices would be needed to analyse the data.

Figure~\ref{fig:angle} shows the correlation of the total width of the
signals with the inclination angle of the track, which behaves as expected.

In principle it would be thinkable, that this feature could be used
to suppress the low angle background already on the chamber level by
cutting on the cluster width.
In figure~\ref{fig:eff2} the efficiency for tracks as a function 
of the inclination angle for different such cuts shows, that this method
in principle works. By choosing a threshold of 7 the efficiency for
tracks with an angle larger than 30$^\circ$ is reduced by a factor of
three without loosing too much efficiency at the nominal inclination
angle of 0$^\circ$. 

However this feature depends so critical on the cut value, that even 
small instabilities in the signal size of the chamber would have big effects
on the background acceptance or even on the efficiency of the nominal
tracks.

\section{Conclusions}

Large triple GEM detectors have shown stable operation in the worst
hadronic beam environment at PSI, giving a large signal gain of
more than $10^4$ with spark probabilities per incoming $\pi$
below $10^{-10}$. 

By using a ``zig-zag'' geometry of the readout board its capacity
could be reduced by a factor three to four, which would proportionally
reduce the thermal noise of the preamplifier, thus allowing to run
the chamber with a gain of about 5000, where the spark probability
is far below $10^{-12}$.

This very high rate capability is supported in addition by an intrinsically
very fast signal response without showing any ``ion tail'' in the detected
signal.

The cluster size is comparable with expectation from the transverse diffusion
of the charge cloud. The FWHM diameter of the signal at the readout board of
about 0.5 mm allows in principle very accurate position measurements with
typical readout line pitches of 0.3 mm, to be measured in a future test beam
setup with high energy particles. But for very high rate application this 
large clustersize can result in significant channel occupancy.

Reducing acceptance of low angle tracks by cutting on the clustersize works
in principle, but the efficiencies are very sensitive to the actual cut value.

\v{0.5}

In summary the triple GEM detector is a very stable, robust, low $X_0$, fast
and cheap detector for applications with intermediate size areas and very
high, even hadronic particle densities, showing lowest sparking rates at very
high gains compared to other technologies. Low angle tracks should be avoided.
Significant safety factors for the gas gain should be used to compensate for
inhomogenous signal size across the chamber. If analog readout is used a very
high accuracy in track position measurement can be expected.


\begin{thebibliography}{99}


\bibitem{GEM} F. Sauli: {\it The Gas Electron Multiplier (GEM)}, Nucl.  Instr.
  and Meth. A 386 (1997) 531-534

\bibitem{triplegem1} M. Ziegler, P. Cwetanski, U. Straumann: {\it A triple GEM
    detector for LHCb}, LHCb public note 99-024, June 30, 1999.

\bibitem{proposal} LHCb technical proposal, CERN LHCC 98-4, 20
  February 1998.
 
%\bibitem{ziegler} M.~Ziegler, P.~Cwentanski, U.~Straumann, LHCb internal note TRAC 99-024 (1999).

\bibitem{talanov} V. Talanov: {\it Radiation environment at the LHCb innner
    tracking area}, LHCb public note 2000-013, May 28, 2000.

\bibitem{perroud} J.P. Perroud, results from a GEM spark test with $\alpha$
    particles, private communication.

\bibitem{bachmann} S.~Bachmann et al.: {\it Charge Amplification and transfer Processes in the Gas Electron Multiplier}, CERN-EP /99-49

\bibitem{helix} HELIX Manual\\ 
http://wwwasic.ihep.uni-heidelberg.de/\~{}feuersta/projects/Helix/helix.ps.gz

\bibitem{lev} L. Shekhtman, Novosibirsk. Private communication in 
  the LHCb inner tracking meeting, CERN, March 17, 2000.


%\bibitem{Sauli} A. Bressan et al. {\it High rate behavior and discharge limits in micro-pattern detectors}, CERN-EP /98-139

%\bibitem{Russen} B. Bochin et al. {\it X-ray tests of double and triple-GEM detectors}, LHCb 98-068 TRAC



\bibitem{cwetanski} P. Cwetanski: {\it Studies on detector prototypes for the
    inner tracking system of LHCb}, Diploma thesis, Universit\"at Heidelberg,
  March 2000.

\bibitem{malte} M. Hildebrandt: {\it Entwicklung und Bau der Detektoren für das 
    Innere Spurkammersystem bei HERA-B}, Ph.D. thesis, Universit\"at Heidelberg,
  April 1999.

\bibitem{alignment} L. Ropelewski, CERN, private communication.

\end{thebibliography}
\end{document}